\documentclass[10pt]{iopart}
\usepackage{iopams}  

\usepackage{graphicx}  
\usepackage{graphicx}  

\def\msun{\,{\rm M_\odot}}

\newcommand\be{\begin{equation}}
\newcommand\ee{\end{equation}}
\newcommand{\ba}{\begin{eqnarray}}
\newcommand{\ea}{\end{eqnarray}}

\begin{document}

\title[Massive black hole cosmic history]{A practical guide to the massive black hole cosmic history}

\author{A Sesana$^1$}

\address{$^1$\ Max-Planck-Institut f\"ur Gravitationsphysik, Albert Einstein Institut, Am M\"ulenber 1, 14476 Golm, Germany}

\begin{abstract}
I review our current understanding of massive black hole (MBH) formation and evolution along the cosmic history. After a brief introductory overview of the relevance of MBHs in the hierarchical structure formation paradigm, I discuss the main viable channels for seed BH formation at high redshift and for their subsequent mass growth and spin evolution. The emerging hierarchical picture, where MBHs grow through merger triggered accretion episodes, acquiring their mass while shining as quasars, is overall robust, but too simplistic to explain the diversity observed in MBH phenomenology. I briefly discuss which future observations will help to shed light on the MBH cosmic history in the near future, paying particular attention to the upcoming gravitational wave window.       
\end{abstract}

\section{Introduction}
There is nowadays a general consensus that massive black holes (MBHs) are fundamental pieces of the puzzle of the galaxy evolution process in the Universe. $10^8-10^9\msun$ MBHs have long been recognized to be the central engines of powerful quasars, whose extreme luminosity ($>10^{45}$erg s$^{-1}$) emitted from a tiny volume ($<10^{-3}$pc, as inferred by variability studies) can be explained only through radiatively efficient accretion of matter onto supermassive compact objects. Observations of quasars out to redshift $\lesssim 7$ \cite{fan01,fan03,mort11} imply that some of these objects were already in place when the Universe was less than a billion year old, although the bulk of quasar population in the Universe comes later, at $z\approx 2$ \cite{richards09}. In the last twenty years, evidence for the existence of MBHs has also been found in nearby quiescent galaxies. The most compelling evidence involves our Milky Way, where accurate dynamical measurements of stellar orbits have revealed a $4\times 10^6\msun$ MBH \cite{ghez05,gillessen09}. Thanks to reverberation mapping \cite{peterson04} as well as stellar and gas kinematics measurements, the masses of $\sim 100$ MBHs in nearby galaxies have been measured (see, e.g., \cite{gultekin09} and references therein), providing compelling evidence that MBHs are ubiquitous in the center of massive nearby galaxies \cite{richstone98}. Their masses have been found to correlate with several properties of the hosting galaxy, in particular with the bulge mass, luminosity and velocity dispersion  \cite{mago98,fer00,geb00,tremaine02,marconi03,haring04,gultekin09,graham11}, and probably with the mass of the dark matter halo (\cite{fer02}, although such relation is debated, see \cite{kor11}), indicating an intimate connection linking the MBHs to their hosts. Such intimate connection is supported by the strong correlation between the redshift evolution of the global star formation rate and the luminosity density of optically selected quasar \cite{boyle98,shankar09}. Moreover, the dustiest starburst galaxies (high redshift submillimeter galaxies and ultraluminous infrared galaxies, ULIRGs) show sign of a recent merger, and are often associated with quasar activity \cite{sanders96,alexander03,swinbank04}. 

Putting the pieces together, a well defined picture has emerged, in which the dormant MBHs that populate galaxies today are the relics of the luminous quasars that shone in the past. In the widely accepted $\Lambda$CDM cosmologies, structure formation proceeds in a hierarchical fashion \cite{wr78}, in which massive galaxies are the result of several merging and accretion events involving smaller building blocks. In this framework, the MBHs we see in today’s galaxies are expected to be the natural end-product of a complex evolutionary path, in which black holes seeded in proto-galaxies at high redshift grow through cosmic history via a sequence of MBH-MBH mergers and accretion episodes \cite{kauffmann00,vhm03,hopkins06}. In this general picture, galaxy mergers trigger copious infall of cold gas in the nucleus of the merger remnant \cite{mihos96}, resulting in vigorous star formation and accretion on the central MBH. The energy output of the accreting MBH acts as a feedback on the surrounding environment either by removing \cite{menci04} or heating \cite{croton06} the cold gas, regulating the star formation process and self-regulating its own accretion \cite{dimatteo05}, possibly resulting in the observed MBH-host relations \cite{silk98,king03}. Hierarchical models for MBH evolution, associating quasar activity to gas-fueled accretion following galaxy mergers, have been successful in reproducing several properties of the observed Universe, such as the present day mass density of nuclear MBHs and the optical and X-ray luminosity functions of quasars \cite{haiman2000,kauffmann00,wyithe2003,vhm03,croton06,malbon07,dimatteo2008}. Though generally successful, most of the key ingredients of the picture are poorly understood. It is still unclear how and when the first seed BHs at high redshift form, and which accretion channel may possibly lead to $>10^9\msun$ MBHs at $z\approx 7$. The MBH-environment interplay establishing the observed MBH-host relation is poorly known, and beyond the resolution of state of the art simulations. Moreover, especially at the low mass end, most of the MBH accretion seems to occur in isolated systems 
, indicating that the merger-accretion paradigm is not the whole story. 
 
A complete understanding of the formation and evolution of MBHs is one of the most exciting goals of contemporary astrophysics and cosmology. Here we review the key ideas composing this fascinating puzzle; the first seed formation and their subsequent mass growth, the galaxy merging process leading to the formation of MBH binaries, the possible paths for the evolution of MBH spins. We caution that, given the vastness of the covered topics, this review is necessarily incomplete. As a matter of fact, each of the following sections would be worthy several reviews in itself. We try to give a general overview, pointing to the relevant literature for in-dept examination by the reader, with the goal of designing a broad background where other specific topics tackled in this volume can be framed. 

The review is organized as follows. In Section 2 we inspect the main routes for seed BH formation at high redshift in the frame of $\Lambda$CDM cosmology. The subsequent accretion and merging history is considered in Section 3, where we describe different possible MBH mass growth paths. Section 4 is devote to spin evolution and to its relation of the MBH merging history and host galaxy properties. We discuss which future observations will help to shed light on the MBH cosmic history in the near future, paying particular attention to the upcoming gravitational wave window, in Section 5, and we briefly summarize in Section 6.
 
\section{The infancy of the giants: seed black hole formation}
\begin{figure}
\centering
\includegraphics[width=4.0in]{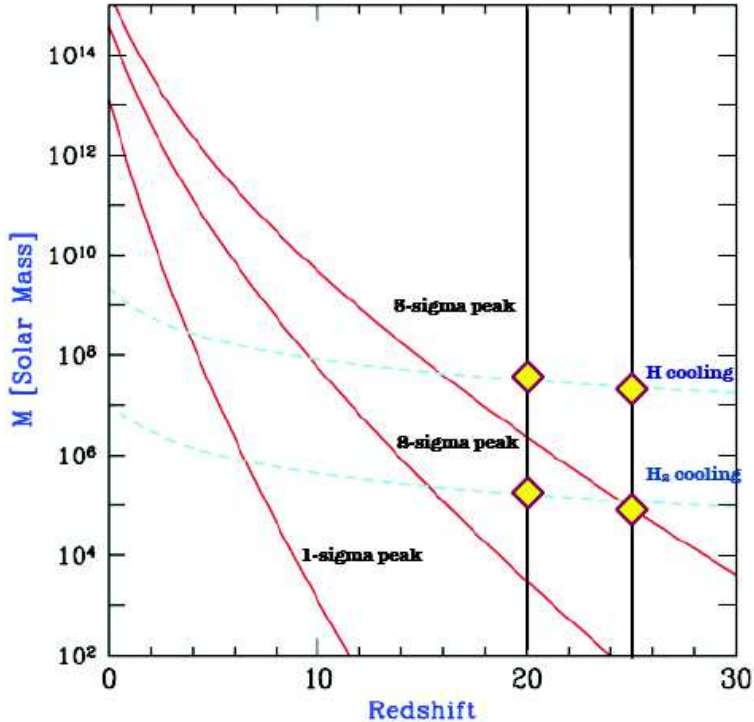}
\caption{First DM halos allowing for efficient gas cooling. Red lines show the mass of different $\sigma$ peaks of the halo mass function as a function of redshift; dashed blue lines show the minimum mass required for efficient H$_2$ (lower) and H (upper) cooling. Diamonds at $z=20$ and $z=25$ mark halo masses needed for efficient H and H$_2$ cooling at those redshifts. Adapted by M. Volonteri from \cite{barkana01}.}
\label{fig1}
\end{figure}

The first obvious relevant question is where and when the seeds of the MBHs powering quasars at redshifts as high as 7, as well as lurking in the center of inactive galaxies today, form. In this section we describe possible paths of seed BH formation in the currently favored general framework of $\Lambda$CDM cosmology. In the standard $\Lambda$CDM picture, the mass content of the Universe is largely dominated by cold dark matter (DM), with baryons contributing to a $10\%$ level only. Starting from a Gaussian density fluctuation field in a quasi-homogeneous Universe, DM perturbations grow in time, to the point they decouple from the Hubble flow and collapse and virialize forming self gravitating DM halos (see \cite{pad93,coles95} for a comprehensive treatment). This is a bottom up path, in the sense that small halos collapse first and get bigger and bigger by accreting from ambient DM and merging with other halos, in a process known as hierarchical clustering. Press and Schechter \cite{press74} provided a statistical description of the halo growth in terms of $\sigma$ overdensities of a Gaussian density field. Figure \ref{fig1}, adapted from \cite{barkana01} by M. Volonteri, illustrates this concept. At any given time (redshift) $1\sigma$ fluctuations of the density field correspond to the characteristic mass of a collapsing halo at that redshift. More massive halos correspond to rarer peaks (i.e., 2-3$\sigma$) of the density field; such halos are much rarer, but they constitute, at each redshift, the densest environments where the formation of the first collapsed objects took place \cite{haiman96,abel02}.

In the early halo assembly process, baryons follow the DM halo potential well, without plying any relevant dynamical role. The first halos suitable to seed BH formation are those exceeding the Jeans limit, $M_J\approx 10^4\msun[(1+z)/10]^{3/2}$, i.e. the mass needed for the self gravity of the halo to overcome thermal gas pressure, allowing the gas to contract together with the halo and to shock heat at the halo virial temperature. Note that $M_J$ is a increasing function of $z$, whereas the characteristic halo mass increases with decreasing $z$. Baryons start to efficiently virialize in rare DM halos of $\sim 10^5\msun$ at $z\approx 50$. The formation of collapsed objects (stars, BHs etc.) within this halos depends on the ability of the gas to cool efficiently. The primordial zero metallicity gas, is mostly composed of atomic and molecular hydrogen \cite{shapiro87}, whose efficient cooling is possible at temperatures higher than $10^4$K and 300K respectively. At any given $z$, the characteristic halo virial temperature of a DM halo is given by $T_v\approx2\times 10^4\,{\rm K}\,M_{h,8}^{2/3}(1+z)/10$, where $M_{h,8}$ is the halo mass in units of $10^8\msun$ \cite{barkana01}. Efficient cooling is therefore possible in rare, high $\sigma$ fluctuations of $\sim 10^6\msun$ at $z\approx30$ for molecular hydrogen, or of $\sim 10^8\msun$ at $z\approx10$ for atomic hydrogen \cite{tegmark97}, as indicated by the intersection of the dotted and solid curves in figure \ref{fig1}. The precise outcome of such cooling process depends on largely unknown details of the involved micro and macro physics, as well as on the evolution of the environment; in the following we spell out the three main seed BH formation mechanisms that has been proposed, the resulting seed mass functions are plotted in figure \ref{figseeds}.

\subsection{Population III star remnants}    
After recombination, a small amount of hydrogen combines in H$_2$ molecules \cite{galli98}. In absence of any primordial soft UV background, H$_2$ is not photodissociated and acts as an effective coolant in $\sim 10^6\msun$ halos at $z\approx30$ with $T_v\gtrsim300$K \cite{tegmark97}. In absence of metals, H$_2$ driven contraction is subsonic, preventing the cloud to fragment (differently to present day metal enriched star forming clouds). Several simulations of the primordial collapse of these metal free clouds suggest that the first generation of stars (the so called PopIII stars) was massive $m_*>100\msun$ (\cite{bromm99,abel00,yoshida06,gao07}, but see discussion below). The final fate of such massive low-low metallicity stars has been extensively studied by Heger and collaborators \cite{heger03}; if $m_*>260\msun$, after only $\sim2$Myr, the star directly collapses into a BH of half its initial mass, leaving behind a few hundred solar mass remnant that can possibly play the role of a seed BH \cite{madau01}.

This very natural scenario presents, however, a lot of uncertainties, mostly related to the mass of the final stars. Early simulations just showed a $\sim 10^3\msun$ gas overdensity surrounding a $0.01\msun$ optically thick core. The mass is then accreted on the stellar core, but the accretion rate and the final mass of the stars are likely affected by feedback effects \cite{mckee08}, possibly resulting in lighter stars. Moreover, recent higher resolution simulations found a significant degree of fragmentation of the collapsing clump \cite{turk09,stacy10}. There are indications that PopIII stars usually form in binary reach clusters with a much less top heavy mass function as previously thought. Typical stellar masses cover a wide $0.1-50\msun$ range \cite{clark11,greif11}, well below the $260\msun$ threshold required for efficient seed BH formation. In principle, a compact cluster of massive stars can still undergo runaway collision, forming a massive single star \cite{pz04}, eventually collapsing in a seed BH, a scenario we will discuss later in Section 2.3. In any case, these recent results are seriously challenging the viability of PopIII remnants as seed BH candidates.

\subsection{Direct collapse}    
Another route to seed BH is via the formation and subsequent direct collapse of a very massive star or disk of gas at high redshift. Several variants of this process have been studied in the literature \cite{haehnelt93,loeb94,eisenstein95,bromm03,koushiappas04,begelman06,lodato06,rossi11}; the advantage with respect to the PopIII scenario is the formation of fairly massive seeds (usually $\sim10^5\msun$), which can more easily grow to $10^9\msun$ to power the observed $z>6$ quasars. 

It has been suggested that direct collapse is efficient only in metal free halos with $T_v\gtrsim10^4$K ($M_h>10^8\msun$), where H$_2$ cooling is suppressed. Under such condition, the gas cloud collapses isothermally at $\sim 8000$K set by atomic line cooling. \cite{bromm03} carried numerical SPH simulations of this scenario, finding a central gas condensation of $>10^6\msun$ in a radius of $<1$pc, without any signal of fragmentation.  However, \cite{dijkstra08} showed that H$_2$ suppression is rather inefficient in such massive halos at this high $z$, and it is likely that H$_2$ also acted as efficient coolant. Moreover, the gas cloud has a significant angular momentum related to its halo spin \cite{bullock01}. For typical halo parameters, the gas forms a rotationally supported disk on $10$pc scales \cite{volonteri10}, and additional angular momentum transport is required to reach the MBH formation conditions. 

Dynamical instabilities, either global or local, provide a viable solution to the angular momentum transport problem. Centrifugally supported clouds become bar unstable if the level of rotation is higher than a given threshold. Bars are efficient in transporting angular momentum outwards creating global inflows of matter. If the gas can cool efficiently, the instability grows on smaller scales, creating a cascade process, called "bars within bars" instability \cite{shlosman89}. Numerical simulations of the process show that it allows accumulation of $>10^6\msun$ of gas at subparsec scales, and it works also in presence of star formation. An alternative way to transport angular momentum is via local instabilities \cite{lodato06}. By acquiring mass from the ambient gas, the disk eventually feels its selfgravity, becoming marginally stable. Disk stability can be described via the Toomre parameter $Q=\frac{c_sk}{\pi G\Sigma}$, where $c_s$ is the sound speed, $k$ the epicyclic frequency and $\Sigma$ the disk surface density. A marginally stable disk has $Q\sim1$. In this situation, rather than fragmenting, the disk develops spiral structures that efficiently transport angular momentum outwards, causing mass inflows in the center \cite{wise08}. Such mass inflows have a double self-regulating effect: the disk surface density gets lower, and the epicyclic frequency of the disk increases, preventing the $Q$ parameter from becoming too small resulting in disk fragmentation. \cite{lodato06} showed that also in this way $\sim10^6\msun$ of gas can be accumulated in the center.

The accumulated gas, will eventually form a massive seed BH. If the mass accumulation timescale is fast enough, a supermassive star (SMS, \cite{hoyle63}) will form in the center. Fully relativistic simulations of isolated rotating SMSs showed that they eventually collapse in a Kerr MBH with negligible mass loss \cite{saijo02,shibata02,montero11}. However, the mass is usually accreted onto the SMS core gradually, and the isolated approximation may be too simplistic. Begelman and collaborators \cite{begelman06} studied in details the SMS formation and evolution. They found that, as accretion proceeds, the central optically thick core collapses into a few $\msun$ BH, accreting efficiently at the Eddington limit of the whole external envelope (called quasistar). The allowed accretion rate therefore greatly exceeds the Eddington rate for the BH only, resulting in its rapid growth. The resulting BH mass is in the $>10^4-10^5\msun$ range. However, recent calculations \cite{rossi11} taking into account for strong wind driven mass loss of the quasistar show that the final mass of the seed BH can be considerably smaller compared to the quasistar mass. In this case, central gas accumulations $>10^7\msun$ are required in order to leave a final seed with mass  $>10^4\msun$, setting a minimal halo mass for efficient seed BH formation of $M_h>10^9\msun$. In such scenario, seed BHs would be rarer and would form at slightly lower redshift ($z\approx12$).
   
\subsection{Runaway stellar dynamics}    
BHs in the $10^2-10^4\msun$ range can be the endproduct of runaway collisions in dense stellar clusters \cite{begelman78,ebisuzaki01,pz02,pz04,gurkan04}. The runaway process has been studied in detail in the context of intermediate MBH formation in globular and massive star clusters. Dense stellar clusters have negative heat capacity, implying that they tend to get denser in the center (a phenomenon known as core collapse \cite{spitzer69,spitzer87}) while the less bound stars get unbound. Core collapse usually occurs on a relaxation timescale $t_{\rm relax}$, but is much faster if the mass spectrum of the involved stars cover a broad range. In this case, the most massive stars segregate to the center in a timescale $t_{cc}\ll t_{\rm relax}$ ($t_{cc}$ stands for core collapse time). If $t_{cc}<3$Myr (the lifetime of the most massive stars), segregated stars undergo runaway collisions forming a very massive star (VMS, \cite{pz99}). This usually happens in clusters more massive than few$\times10^5\msun$ with half mass radius $<1$pc. Devecchi and Volonteri \cite{devecchi09} showed that runaway collapse is a viable route for seed BH formation in mildly metal polluted ($Z<10^{-3}Z_\odot$), efficient hydrogen cooling ($T_v>10^4$K) halos at high redshift. Their scenario is similar to the direct collapse mechanism discussed above, with the difference that the mass accumulated in the central parsec overcome the density threshold (function of the metallicity, see \cite{santoro06}) at which fragmentation and star formation occur efficiently. The most massive stars in the resulting compact cluster undergo runaway collisions forming a VMS of several thousand solar masses, leaving behind a seed BH remnant of $\sim 1000-2000\msun$. As already mentioned, a similar process may operate at zero metallicity even during the first episode of star formation: if PopIII stars form in clusters, following fragmentation of the collapsing cloud, runaway collisions can in principle still lead to the formation of a  $\sim 1000\msun$ seed BH. Davies and collaborators \cite{davies11} recently proposed a variant of this scenario, in which the requirement $t_{cc}<3$Myr is relaxed. If a dense star cluster is formed, subsequent massive gas inflows following frequent halo mergers at high redshift cause significant contraction of the cluster core, which undergoes a global collapse. Stellar remnants merge together leaving behind a large seed BH of $\sim10^5\msun$, comparable to direct collapse endproducts.   

   \begin{figure}
   \centering
   \includegraphics[width=4.0in]{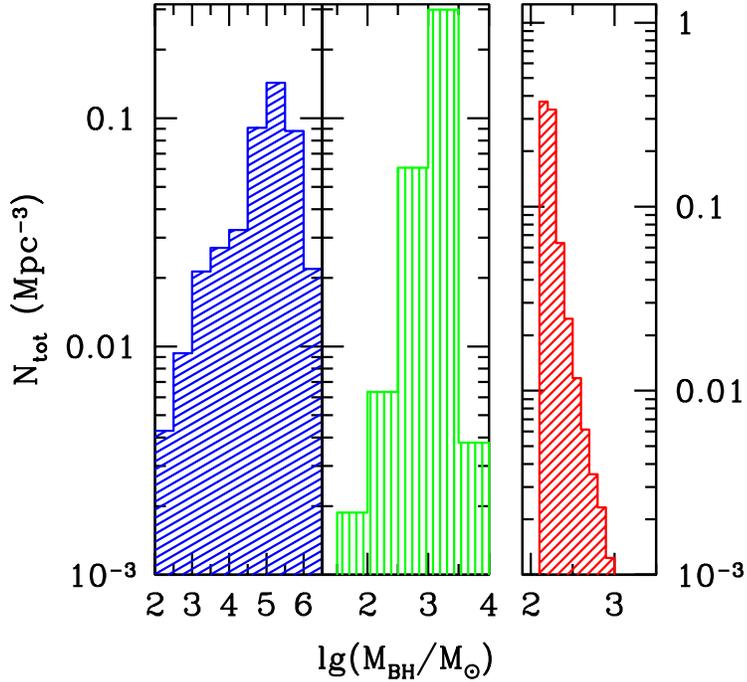}
      \caption{Mass function of seed BHs for the three different formation scenarios discussed in the text: direct collapse (left panel), runaway mergers in high redshift star clusters (central panel), and PopIII remnant (right panel). From \cite{volonteri10}.}
         \label{figseeds}
   \end{figure}

\section{The making of the giants: massive black hole accretion history}
Once they formed, seed BHs will inevitably interact with their environment. Nearby elliptical galaxies host MBHs of $>10^9\msun$, and observations of distant quasars imply that such monsters (at least few of them) where already in place at redshift $\lesssim 7$. Therefore, seed BHs need to accrete an enormous amount of mass, and they need to do it fast! We can identify three principal mass growth mechanisms:
\begin{itemize}
\item merger with other MBHs; 
\item episodic accretion of compact objects, disrupted stars, or gas clouds;
\item prolonged accretion of large supplies of gas via accretion disks. 
\end{itemize}
Soltan \cite{soltan82} first noticed that the optical luminosity function of quasars directly implies a large population of nuclear MBHs lurking in quiescent galaxies today. Infact, to an observed luminosity $L$ corresponds a mass accretion rate $\dot{M}=f_{\rm bol}(1-\epsilon) L/(\epsilon c^2)$, where $f_{\rm bol}$ is a bolometric correction to the luminosity and $\epsilon$ is a mass-to-energy conversion efficiency. Subsequent mass measurements of the local nuclear MBHs suggested that the MBH mass density in the local Universe is consistent with the accreted mass inferred by integrating the quasar luminosity function at all redshifts assuming $\epsilon\approx0.1$ \cite{yu02,elvis02,marconi04,shankar04,shankar09} . Therefore, the 'quasar mode', in which large amount of gas are accreted in single coherent episodes via accretion disks shining close to the Eddington luminosity \footnote{We recall that the Eddington luminosity is the maximum admitted luminosity for which the radiation pressure exerted the photons emitted in the accretion process is smaller than the gravitational binding energy of the accreting material. If the contrary is true, radiation pressure blows away the reservoir of gas, suppressing the accretion process. For standard radiatively efficient accretion flows \cite{shakura73}, the Eddington luminosity corresponds to an accretion rate of $\dot{M}_{\rm Edd}\approx2.5 M_8\msun$yr$^{-1}$, where $M_8$ is the MBH mass normalized to 10$^8\msun$.}, appears to be principal path of MBH cosmic growth. We notice, however, that there are caveats to this argument. If $z\approx 2$ quasars have $\epsilon\approx0.2$ (which may be the case if they have, on average, a substantial spin), then the integrated MBH mass density implied by the quasar luminosity function would be much lower, leaving room to a significant contribution from obscured accretion. Infact, hard X-ray source counts suggest that most of the MBH growth occurs in highly obscured objects (\cite{bauer04,fiore08}, see Treister and Urry contribution to this volume). A significant 'obscured' growth is still consistent with the Soltan argument (that is usually applied to the optical quasar luminosity function) if the typical $\epsilon$ is high. We notice, moreover, that the local MBH mass density has in itself a factor of two uncertainty, readily allowing equal amount of obscured and unobscured accretion even assuming $\epsilon\approx0.1$. 

The Soltan argument is consistent with the general picture in which galaxy mergers trigger gas inflow in the nuclear region, feeding quasar activity \cite{kauffmann00,hopkins06,ellison11}. A $10^9\msun$ MBH is thought to have undergone several of such mergers in its lifetime, acquiring most of its mass through relatively short ($\sim 10^7$yrs) Eddington limited accretion episodes. Although this general picture has proved successful in reproducing the properties of the observed quasar population (see, e.g., \cite{vhm03,malbon07}), the detailed accretion history of MBHs is far more complex. Firstly, it is hard to grow $10^9\msun$ MBHs at redshift $\approx7$, when the Universe was $\sim0.8$Gyr old; it is therefore likely that early accretion at high redshift proceeded at super-Eddington rates (see next section). Secondly, although the merger driven paradigm seems to apply to bright quasars, the same is not true for fainter, less massive, accreting MBHs. Most of those are hosted in spiral galaxies that likely did not undergo any major merger in their lifetime. There is a wide variety of accretion modes, likely related to the environmental conditions surrounding the MBHs, and all of them have to be explored in order to get a comprehensive picture of the MBH accretion history.

\subsection{The highest redshift quasars}
One of the main challenges of hierarchical formation models is to address the presence of $>10^9\msun$ MBHs at $z\lesssim7$, when the Universe was less than 0.8Gyr old. Note that these systems have a space densities of 1 Gpc$^{-3}$ comoving volume \cite{fan03}, corresponding to $5\sigma$ peaks fluctuations of the original density fields \cite{haiman01}. This means that they are not representative of the typical MBH cosmic evolution; they are extremely rare, and grew in the richest overdensities of the early Universe. To grow to $>10^9\msun$ at $z\lesssim7$, a putative seed BH would require almost continuous Eddington limited accretion over its whole lifetime. The mass accretion depends both on environmental factors (the supply of available gas) and the detailed accretion process (e.g., the effects of accretion feedback). The environment does not seem to be a problem for these objects. Numerical simulations show that the nuclear regions of extreme high-$z$ overdensities are constantly fed by filamentary cold gas flows \cite{dekel06,dekel09}. If efficiently accreted, the MBH can grow to $>10^9\msun$ without even experiencing a merger event with another protogalaxy hosting another MBH. Therefore, the merger triggered quasar accretion scheme seems to break down in such extremely dense environments.    

Whether the mass supplied through the filaments can be efficiently accreted is a delicate issue. Eddington limited accretion for $<10^9$yrs can produce a MBH of $>10^9\msun$ only if $\epsilon\lesssim0.1$, i.e. if the MBH spin remains small \cite{yoo04,li07}. Infact, writing the mass growth as $M(t)=M(0){\rm exp}\left(\frac{1-\epsilon}{\epsilon}\frac{t}{t_{\rm Edd}}\right)$, where $t_{\rm Edd}=0.45$Gyr, is easy to see that it needs $\approx0.5$Gyr to grow a 10$^9\msun$ MBH out of a PopIII seed if $\epsilon=0.06$ (accretion onto a Schwarschild BH), but it takes $\approx5$Gyr if $\epsilon=0.4$ (prograde accretion onto an highly spinning BH). However, continuous coherent accretion through a standard accretion disk \cite{shakura73} is expected to efficiently spin-up the MBH \cite{bardeen70,thorne74}, increasing $\epsilon$ to $>0.3$, hence considerably slowing down the MBH growth. To solve this problem King and collaborators \cite{king05} proposed that mass is accreted in a series of small incoherent packets (chaotic accretion). In this case, depending on the angular momentum of the accreted material, the MBH is spun up or down, performing a random walk in spin magnitude that keep it close to zero \cite{volonteri07}. Although such solution is appealing (and probably relevant in different contexts, see Section 3.3), a continuous supply of gas powering a quasar for such long time ($\sim10^9$yr) is likely to be coherent, settling in a well defined accretion disk, implying efficient spin-up.

It has been suggested that MBH growth at high redshifts proceeds through radiatively inefficient accretion flows. If the mass supply rate is much larger than the Eddington rate ($\dot{m}=\dot{M}/\dot{M_{\rm Edd}}\gg1$, which can be easily sustained by cold inflows at high $z$), photons are trapped in the accretion flows because the time it takes them to diffuse out of the flow is longer than the accretion time \cite{begelman78b,begelman82}. Consequently, the emitted luminosity is suppressed to the point that $\dot{m}$ can greatly exceed 1 (supercritical accretion), still resulting in a sub-Eddington luminosity. Note that early supercritical accretion is still consistent with the Soltan \cite{soltan82} argument: MBHs can in principle grow this way to $10^7-10^8\msun$, complete the last couple of e-folding in mass via radiatively efficient accretion, and still satisfy the quasar luminosity function-mass conversion constraints. However, simulations of radiative inefficient accretion flows \cite{stone99,hawley02,igumenshchev03,proga03,ohsuga05} showed that most of the infalling matter is driven away by wind-like outflows, and the actual accretion rate can exceed the Eddington limit by a factor of $\approx 10$ only. Supercritical accretion is also allowed in quasi-spherical Bondi-like accretion flows \cite{begelman78b}. Volonteri and Rees \cite{volonteri05} discussed how a short phase of supercritical accretion can easily produce $>10^9\msun$ MBH at $z>6$ out of PopIII remnant seeds, however how and when this happens in practice is an open question.    
      
\subsection{The standard paradigm of MBH evolution}
Most of the MBH mass density of the Universe is built-up during the peak of quasar activity at $z\approx2$. At those low redshifts, MBHs had all the time to grow in mass through a sequence of merger-triggered, Eddington-limited accretion events, without invoking continuous high redshift cold gas inflows or supercritical accretion. As already discussed, standard hierarchical MBH formation models have been proved successful in producing the population of MBHs powering medium redshift quasars, and their luminosity function. 

One of the main features of the hierarchical assembly, is the formation of a large number of MBH binaries along the cosmic history. A detailed description of their dynamical formation and evolution can be found in Dotti et al. contribution to this volume. In few words, following galaxy mergers, MBHs sink to the center of the remnant owe to dynamical friction against the DM/stellar/gaseous background \cite{colpi99,ostriker99}. At subparsec scales, the mass enclosed within the relative orbits of the two MBHs becomes less then their own mass; the MBHs feel each other pull forming a bound binary. The binary fate is defined by its interaction with the stellar and gaseous environment. In gas rich nuclei, torques exerted by a circumbinary disk can effectively extract the MBH binary energy and angular momentum, possibly resulting in a fast ($\sim10^7$yr) shrink to milliparsec separations \cite{escala05,dotti07}, where gravitational wave (GW) emission efficiently drives its coalescence \cite{peters64}. Also in gas poor environments, recent simulations \cite{berzcik06,kahn11,preto11,gualandris11} showed that three body interactions with ambient stars efficiently fed (by rotation and triaxiality of the stellar distribution) into the binary loss cone can bring the system to final coalescence in $\sim10^8$ years. Along with the binary formation, during the merger, the cold gas content of the interacting system is highly destabilized, triggering inflows in the nuclear region \cite{barnes96,mihos96} that provide a large reservoir of fuel for the active phase of the MBHs. Concurrently, the dense nuclear cold gas trigger an efficient star formation episode \cite{hopkins06}, which competes with the MBH gas fueling depleting the reservoir of available gas. MBH feedback onto the surrounding environment regulates the MBH growth, shaping the MBH-bulge relations observed in the local Universe. This general scenario naturally reproduces the observed 'downsizing' \cite{cowie96,cowie03,ueda03}, which was initially thought to be in contrast with the hierarchical formation process. The downsizing is the phenomenon whereby luminous activity appears to occur in progressively lower mass objects as the redshift decreases; i.e., the most massive MBHs seem to accumulate their mass before the lighter ones \cite{merloni04}. As noted by Malbon and collaborators \cite{malbon07}, downsizing is a natural outcome of hierarchical structure formation. Even without invoking complex feedback mechanisms, the most massive black holes are those that grew in the densest environments, where dynamical evolution occurs on shorter timescales. Such MBHs are likely to experience several early mergers that led to an early exhaustion of their cold gas reservoir (because of accretion, star formation and feedback), suppressing their luminous activity at lower redshift. Conversely, less massive objects are hosted in less dense environment and experience a smoother evolution: at low redshift is still plenty of cold gas for them to shine as luminous quasars, creating the 'downsizing' effect.  

There is plenty of compelling evidence supporting, to some level, this scenario. The observed star formation history and the quasar activity in the Universe mimic each other peaking at $z\approx 2$ \cite{boyle98,shankar09}; many luminous quasars are found in 'disturbed' galaxies, where the presence of tails or clumps in the matter distribution indicate a recent merger with another object (see, e.g., \cite{hopkins07,urrutia08}); dual quasars have been found in interacting galaxy pairs with projected separation $<50$kpc (NGC 6240 \cite{komossa03} being the prototypical case); most ULIRGs (i.e., highly star forming galaxies) are associated to merging systems and show active galactic nuclei (AGN) in their center \cite{sanders96}. Nevertheless, the overall picture suffers of many uncertainties. Firstly, galaxy mergers generally trigger AGN activity, but the contrary is not true; most of the AGN (especially in the local universe) occurs in 'unperturbed', isolated galaxies (see next Section). Secondly, the relation between the AGN and star formation triggering timescale and the MBH binary evolution timescale is poorly constrained. The presence of dual quasars indicates that at least in some cases accretion is triggered well before merger, however that might not be the general case. Finally, evidence for bound MBH binaries remain circumstantial \cite{valtonen08,komossa08,boroson09,tsalmantza11,eracleous11}, and most of the proposed candidates have different explanations (see Dotti et al. contribution to this volume for a detailed discussion).

\subsection{Secular evolution at the small end of the MBH mass function}
There is growing observational evidence that the merger driven accretion paradigm is not the whole story. In the clustering scenario, elliptical galaxies form as a consequence of galaxy mergers \cite{dimatteo05}, and we can infer that most of the residing MBH mass has been acquired by merger triggered accretion episodes (consistent with the cosmic population of bright quasars). However, present day spirals (which dominate the population of low mass galaxies \cite{bell03}) have likely experienced a much quieter cosmic evolution, possibly without undergoing any major merger event. Still, many of them show significant nuclear activity, shining as Seyfert galaxies \cite{ho97}. Seyfert galaxies are in general characterized by lumpy structures rich of cold gas \cite{pogge02}, and most of them show dense nuclear stellar clusters with densities of $\sim 10^6\msun/{\rm pc}^3$ (an environment similar to the Milky Way center \cite{schoedel07}). In such a rich environment, the main MBH growth channel is thought to be the accretion of small packets of material reaching the galactic nucleus because of dynamical relaxation and secular evolution processes. 

In dense stellar environments, tidal disruptions of main sequence stars scattered in the MBH loss cone by two body relaxation are quite common. During a disruption event, half of the debris is accreted by the MBH, powering a luminous episode which can last for years, with a characteristic power law decline $L\propto t^{-5/3}$ \cite{rees88}. Theoretical studies predict rates in the range $10^{-5}-{\rm few}\times10^{-4}$yr$^{-1}$ in relaxed dense nuclei hosting a $10^6-10^7\msun$ MBH \cite{magorrian99}. Moreover, such rates can be highly increased for a short time ($\sim$Myr) by the presence of a secondary inspiralling MBH as a consequence of a minor merger \cite{chen09,chen11,wegg11} or a star cluster accretion event \cite{pz02b}. To date, a total of $\sim15$ tidal disruption flare candidates has been identified; most of them in X-ray and UV \cite{halpern04,komossa04,gezari06,esquej09,gezari09}, and two in optical \cite{vanvelzen10}. The soft X emission is generally found to be consistent with thermal emission from a $T\sim10^5$K accretion disk, in line with expectations for a $10^6-10^7\msun$ MBH \cite{strubbe09,lodato10}. The inferred rates are broadly consistent with theoretical predictions, implying that this may indeed be a major mass growth channel for MBHs with $M<{\rm few}10^6\msun$ \cite{milosavljevic06}. The presence of segregated nuclear clusters of compact objects \cite{hopman06,preto10} is also expected to result in a significant number of 'dark' accretion events (extreme mass ratio inspirals, EMRIs). Even though the event rates are likely too small to significantly contribute to the MBH mass build-up, such events may offer a unique way to probe MBHs in quiescent nuclei through GW observations (see Section 5).

The principal growth channel of MBH in spirals, however, is likely to be the random accretion of molecular clouds \cite{hopkins06b}. Observations suggest that typical accretion episodes in Seyfert lasts $10^4-10^5$yr, whereas the total active lifetime (based on the fraction of disk galaxies that are Seyfert) is in the range $10^8-10^9$yr \cite{ho97}. Disk galaxies show a rich structure of nuclear molecular clouds down to parsec scales (see, as a reference, the MW \cite{miyazaki00,oka01}), with masses in the range $10^3-10^5\msun$. Relaxation processes will likely drive some of these clouds toward the nucleus, fueling the dormant MBH. This feeding channel is likely unimportant in giant ellipticals, where the nuclear densities are much lower, the bulge less lumpy, and the content of cold gas very small \cite{sage07} as a consequence of AGN feedback and stellar winds loss related to the last accretion episode (see, e.g., \cite{ciotti07}), resulting in a pronounced lack of molecular clouds.

\section{Spin evolution}
Astrophysical black holes are fully described by two quantities only: mass and spin. The latter can be expressed by the dimensionless parameter $\hat{a}=J/J_{\rm max}=cJ/GM_{\rm BH}^2$. By definition $0\leq\hat{a}\leq 1$. At present, we have several effective ways to measure MBH masses, and we can construct MBH mass functions and study their evolution with redshift; conversely, spin measurements are way more problematic, and only a handful of estimates are available at the time of writing. The principal measurement technique is through fitting the skewed relativistic K$\alpha$ fluorescent line \cite{laor91}, whose shape is highly spin-dependent. There is just a handful of objects with reliable spin measurements. Notable examples are: 
Fairall 9, $\hat{a}=0.60\pm0.07$ \cite{schmoll09} and $\hat{a}=0.67^{+0.11}_{-0.10}$ \cite{patrick11}; 
1H 0707-495, $\hat{a}>0.93$ \cite{delacalle10};
MRK509, $\hat{a}=0.78^{+0.04}_{-0.03}$ \cite{delacalle10};  
MRK 79, $\hat{a}=0.7\pm0.1$ \cite{gallo11}. However, several free parameters enter in the fitting procedure, creating a severe degeneracy problem. The measurement is particularly sensitive to the modeling of the soft X excess of the continuum, which blends with the low energy tail of the line (that is crucial for the spin estimation). Different fits to the K$\alpha$ line in MCG-6-30-15 gave $\hat{a}>0.98$ \cite{brenneman06}, $\hat{a}=0.86\pm0.01$ \cite{delacalle10} and $\hat{a}=0.49^{+0.20}_{-0.12}$ \cite{patrick11}. Even more problematic is the case of NGC 3783, for which different groups found $\hat{a}>0.88$ \cite{brenneman11} and $\hat{a}<0.32$ \cite{patrick11}.

Measuring and understanding spins is crucial to assess the MBH cosmic evolution. Firstly, spins affect the accretion-luminosity conversion efficiency; highly spinning BHs can convert up to $\sim 40\%$ of the accreted matter in radiation, growing much slowly. Secondly, the 'spin paradigm' for MBH jets \cite{blandford90} assumes that radio jets observed in AGNs are launched by highly spinning BHs \cite{blandford77}. Lastly, spins dramatically affect the gravitational recoil suffered by the remnant MBH after a binary merger. It has infact been shown that highly spinning BHs can experience kicks up to 5000 km s$^{-1}$ depending on their progenitor spin magnitude and orientation \cite{campanelli07,marronetti07,lousto11}. These 'superkicks' are sufficient to eject the remnant from the deepest potential well of the most massive galaxy clusters, with potentially important implications for the hierarchical MBH formation paradigm, and for the occupation fraction of MBHs in galaxies.

\subsection{Black hole spin evolution channels} 
MBH spins can significantly evolve via two major channels: accretion of matter, and coalescences with other MBHs. Different evolution paths lead to different spin distributions, as it has been extensively discussed by Volonteri and Berti \cite{berti08}.
\subsubsection{Coalescences.}
MBH binary coalescence is an unavoidable key element in the hierarchical formation scenario. The binary system is characterized by an orbital angular momentum and two spin vectors, whose sum (plus the angular momentum radiated in GWs) has to be conserved during the coalescence. Under this basic assumption, Hughes and Blandford \cite{hughes03} found that, in general, coalescences of MBHs with random spin directions result in a broad remnant spin distribution; in particular highly spinning MBHs tend to spin-down. With the advent of numerical relativity \cite{pretorius05,baker06,campanelli06}, it is now possible to construct accurate mappings of initial-to-final BH spins following purely GW driven coalescences \cite{rezzolla08}. In general, unless the two initial MBH spins and the binary orbital angular momentum are aligned, coalescences are unlikely to result in highly spinning MBHs.
\subsubsection{Accretion.} 
\begin{figure}
   \centering
   \includegraphics[width=4.0in]{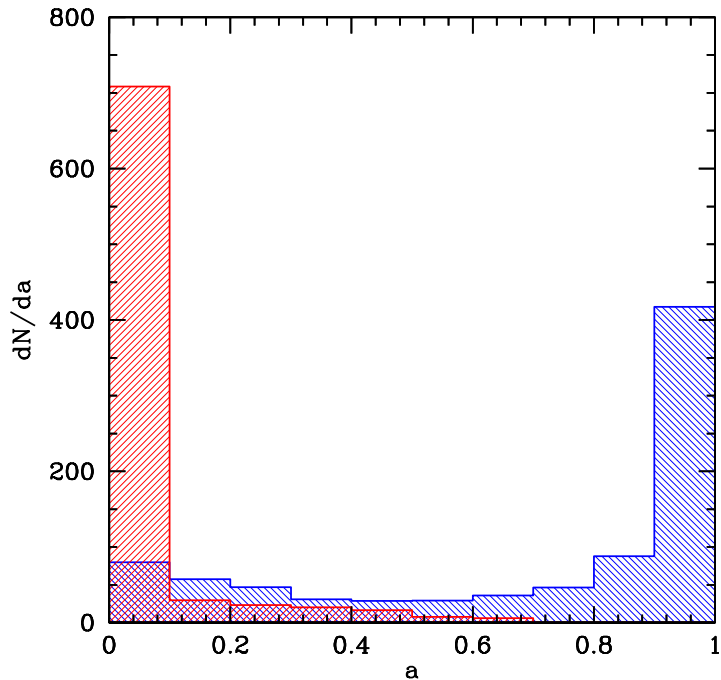}
      \caption{Spin distribution of MBH coalescence remnants for two different MBH accretion models. Red histogram: accretion proceeds in a chaotic fashion according to \cite{king05}, resulting in small spins; blue histogram: accretion is coherent, resulting in efficient spin-up. The seed BH spins at formation were set to zero.}
         \label{figspin}
   \end{figure}

The accretion imprint on the MBH spin crucially depends on the characteristic of the accretion flow. Firstly, let us consider a MBH accreting from an extended planar accretion disk with $M_{\rm disk}\gg M_{\rm BH}$ and $L_{\rm disk}\gg S_{\rm BH}$, where $L_{\rm disk}$ is the disk angular momentum and $S_{\rm BH}$ is the MBH spin. The natural consequence of such accretion mode is MBH spin-up. Given an initial mass $M_0$, coherent accretion of a $\sqrt{6}M_0$ mass of gas will turn a non rotating MBH into maximally spinning \cite{thorne74}. If $L_{\rm disk}$ and $S_{\rm BH}$ are misaligned, the Bardeen-Petterson effect \cite{bardeen75} will act to align the MBH spin to the disk angular momentum in a very short timescale ($t_{\rm align}\ll t_{\rm acc}$ \cite{scheuer96,natarajan98}), leaving the picture unchanged. However, things can be significantly different if $L_{\rm disk} \lesssim S_{\rm BH}$. Defining $\theta$ to be the angle between $L_{\rm disk}$ and $S_{\rm BH}$, King and collaborators \cite{king05} found that if cos$\theta>-L_{\rm disk}/2S_{\rm BH}$ the accreted material counteraligns with the MBH. For $L_{\rm disk} \ll S_{\rm BH}$, there is equal chance of prograde and retrograde accretion; the latter, however, transfers more angular momentum per unit mass, because the counterrotating innermost last stable orbit is larger. If, therefore, MBH growth occurs through accretion of incoherent packets of material with $\delta m_{\rm acc}\ll M_{\rm BH}$ and random angular momentum orientation, their final spins will be generally close to zero (see figure \ref{figspin}).

\subsection{Connecting MBH spins to their accretion history}
Following our discussion on the MBH mass growth in Section 3, the cosmic MBH spin distribution might therefore be multimodal, and strongly correlated with the merging and accretion history (and therefore morphology) of the host galaxies \cite{volonteri07}. In particular MBHs residing in giant ellipticals that experienced several gas rich mergers, are expected to be highly spinning. During such mergers infact, the two sinking MBH spins aligns quickly \cite{perego09,dotti10} to the nuclear disk angular momentum, and both coherent accretion and the final coalescence contribute to the remnant MBH spin-up. Alignment has also important consequences for the gravitation recoil; since binaries with aligned spins experience the weakest kicks, their coalescence remnants will be generally unable to escape the galaxy core. For dry mergers, however, no spin alignment is expected, and the MBH remnants will generally have a broader spin distribution, experiencing higher kicks that can in principle evacuate them from their host galaxies. Many disk galaxies, conversely, may have not experienced any major merger event at low redshift. In this case, if the central MBH spin was initially high, random (chaotic) accretion of molecular clouds or tidally disrupted stars would bring it to low (possibly close to zero) values \cite{volonteri07}. However, we should mention that, in many cases, studies of the kinematic in the nuclear region of spirals have revealed disk like structures in the gas distribution (see, e.g., \cite{davies04a,davies04b}), leaving the possibility of substantial coherent accretion open. Although the question of accretion in spirals is not an neat as in ellipticals, it is likely that the average spins of the MBHs hosted by the formers is lower. This general picture predicts high spins in bright quasars (and present giant ellipticals). If this is the case, and we assume the 'spin paradigm' as the source of radio jets, then elliptical galaxies have to be generally radio louder than spirals, in agreement with the finding of Sikora and collaborators \cite{sikora07}. This scenario is supported also by Capetti and Balmaverde \cite{capetti06} that found that radio bimodality correlates with bimodality of stellar brightness profiles in galactic nuclei. The inner regions of radio-loud galaxies display shallow cores, whereas radio quiet galaxies, including nearby low-luminosity Seyferts, have instead power-law brightness profiles (cusps) and preferentially reside in S0 and spiral galaxies. However, in the modeling of radio loud/radio quiet sources, retrograde accretion is possibly a key ingredient \cite{garofalo09a,garofalo09b} (see \cite{dotti10b} and reference therein for a captivating scenario connecting radio loudness to retrograde accretion). On the other hand, at the time of writing, the paucity (and uncertainty) of reliable spin measurements, does not offer a robust guideline for theoretical modeling. 

\section{Probing massive black holes in the era of gravitational wave observations}
The MBH evolution picture emerging from the previous Sections is rich and diverse, and only loosely bound by observations. Different MBH seeding models and accretion recipes are consistent with current observational constraints about the present day MBH mass density and mass function, the quasar optical and X-ray luminosity functions, the count of faint X-ray sources in deep fields and the unresolved X-ray background \cite{volonteri06,salvaterra07}. On the other hand, the MBH spin distribution is basically unconstrained by observations; just a handful of spin measurement are available, and conjectured spin distributions rely on theoretical models (e.g. the 'spin paradigm' for jet production can be used to infer that MBHs in bright quasars are close to maximally spinning \cite{nemmen07}). 

Several future observations can help placing better constrains to current MBH evolution models. For example, \cite{volonteri09} argues that the imprint of the seed BH population should be particularly evident at the small end of the MBH mass function. We therefore need to target the nuclei of dwarf galaxies, to unveil their central MBH. Depending on the seed formation mechanism, such MBHs might systematically lie above or below the $M-\sigma$ relation observed for masses $>10^6\msun$. Moreover, the MBH occupation fraction in such small galaxies may offer valuable insights on the seed BH formation efficiency \cite{volonteri08} and on the effectiveness of gravitational recoils, ultimately resulting in useful information on the MBH spins. Counting high redshift faint objects in future deep fields with the JWST will also help constraining the evolution of the MBH population at high redshift \cite{volonteribegelman10}, while long exposures with the X-ray telescope Athena \footnote{http://www.mpe.mpg.de/athena/home.php?lang=en} will likely increase the number of measured spins. 

   \begin{figure}
   \centering
   \includegraphics[width=4.0in]{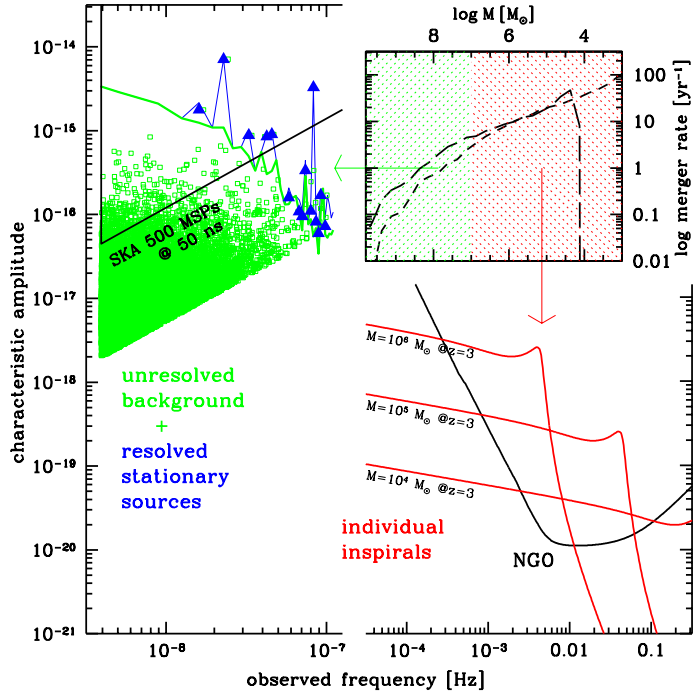}
      \caption{The GW landscape. A representative summary of GW observations of MBH binaries with PTAs (through the detection of a GW stochastic background) and NGO (via the direct detection the final stage of the coalescence of individual binary systems). In the main panel we show the characteristic amplitude of GW signals as a function of frequency, compared to the sensitivity of SKA and NGO. On the left side, we show a Montecarlo realization of the GW signal coming from a cosmological population of MBH binaries: green dots represent the strain of each individual source, the overall signal is shown as a blue solid line, individually resolvable sources are shown as blue triangles, and unresolved background is shown as a green solid line. On the right side, red lines track the inspiral of selected MBH binaries at $z=3$. The upper right panel shows the massive black hole binary coalescence rate as a function of the total mass $M$ for two representative MBH population models. The green and red shaded areas highlight the portion of the mass function that will be probed by PTA and NGO respectively, highlighting the complementarity of the two probes.}
         \label{figlisaska}
   \end{figure}

The next two decades, will witness the dawn of GW astronomy. While signals coming from compact stars and binaries fall in the observational domain of operating and planned ground based interferometers (such as LIGO, VIRGO, and the proposed Einstein Telescope (ET)), MBH binaries are expected to be among the primary actors on the upcoming low frequency stage. The $10^{-4}-10^{-1}$Hz window is going to be probed by the spaceborne New Gravitational Observatory (NGO \footnote{Following the termination of the ESA-NASA partnership at the basis of the Laser Interferometer Space Antenna (LISA) project, a new design study is being submitted to ESA, for a smaller mission called NGO. More information can be found at https://lisa-light.aei.mpg.de/bin/view}), which is under consideration by ESA as the possible next L1 space mission; whereas the nHz ($10^{-9}-10^{-7}$Hz) range will be cover by precise timing of an ensemble of millisecond pulsars (forming a so called pulsar timing array, PTA \cite{manchester08,janssen08,jenet09,lazio09,hobbs10}). While NGO will detect $10^3-10^7\msun$ coalescing MBH binaries, mostly at $z>2$, PTAs will be sensitive to the inspiral phase of genuine supermassive systems ($M>10^8\msun$) at lower redshift ($z<1$), providing a complementary probe of the MBH population in the Universe, as shown in figure \ref{figlisaska}. 

NGO will provide a unique survey of coalescing MBH binaries up to $z\approx 15$ \cite{haehnelt94,hughes02,loeb03,sesana04,enoki04,rook05,sesana05,sesana07} \footnote{All the relevant literature was written considering LISA as reference mission concept; numbers and rates are similar for NGO.}, offering the unique possibility of disentangling different seed formation mechanisms \cite{sesana11a,gair11,plowman11}. Masses and spins of coalescing MBHs will be measured with unprecedented precision (see, e.g., \cite{vecchio04,lang06,lang08,mcwilliams11}), placing strong constrains on the MBH accretion history. Moreover, NGO will also give the opportunity of detecting several (up to a hundred, but the rates are highly uncertain \cite{gair04,amaro07}) extreme mass ratio inspirals (EMRIs), i.e., inspirals of stellar mass compact objects into $10^5-10^6\msun$ MBH, out to $z\lesssim 0.7$. This is particularly important, not only for the unique opportunity to test GR in the strong field regime \cite{barack07}, but also because it will allow to unveil the low end of the mass function \cite{gair10} and the spin distribution (both MBH mass and spin will be measured with an accuracy of $<1\%$ from GW observations \cite{barack04}) of dormant MBHs. Infact, all the MBH observations beyond the local Universe are inevitably biased toward active systems; conversely, EMRIs are likely to occur in dense relaxed nuclei (e.g., the Milky Way) with no preference for the active ones. 

PTAs exploit the characteristic fingerprint left by passing GWs in the time of arrival of the radio pulses propagating from the pulsar to the receiver on Earth \cite{saz78,det79,ber83,hel83,jen05}. Given the low frequency window ($10^{-9}-10^{-7}$Hz), PTAs will be sensitive to the collective signal coming from the inspiralling population of supermassive, low redshift binaries \cite{sesana08,sesana09}, providing a strong test for the effectiveness of the galaxy merger process at low redshift, and helping placing constrains to the high mass end of the MBH mass function. \cite{kocsis11} found that several hundred sources will contribute to the signal at a 1ns level, considered the ultimate goal for the Sky Kilometer Array (SKA \cite{lazio09}). Interestingly, such systems are far from coalescence, and they can still retain much of their original eccentricity against GW circularization \cite{sesana10,preto11,roedig11}. Eccentricity measurement of individually resolved sources may help in constraining the evolution of MBH binaries, testing our current model of their dynamical evolution in star/gas dominated environments. Moreover, the identification of putative electromagnetic counterparts will open new avenues in the era of multimessenger astronomy \cite{sesana11b,tanaka11}.    

\section{Summary}
The cosmic evolution of MBHs is one of the most diverse and fascinating puzzles of the structure formation in the Universe. MBHs are ubiquitous in galactic nuclei today, and their tight correlations with their hosts witness an intimate link between their evolution and galaxy formation. In this review, we followed this link reconstructing the key elements of the puzzle. It is now widely accepted that the seeds of the MBHs powering quasars and lurking in the galactic centers today were formed in protogalactic DM halos of $10^6-10^8\msun$ at high redshift, $z\approx20$. The details of the formation process are largely unknown, and at least three distinctive mechanisms have been put forward: (i) light ($\sim100\msun$) seed formation as PopIII star remnants; (ii) massive ($\sim10^5\msun$) seed formation from direct collapse of protogalactic massive gas clouds; (iii) intermediate ($\sim10^3\msun$) seed formation following runaway dynamics in protogalactic star clusters. The subsequent seed growth is largely determined by the environmental conditions in which their formed. Seeds populating the deepest overdensities in the early Universe are likely to grow fast, possibly at a supercritical rate, fed by massive streams of cold gas, forming the rare massive objects powering the $z\approx7$ quasars we see today. The MBHs powering the much more common $z\approx2$ quasars may instead have experienced a less violent fate, undergoing a series of merger events triggering Eddington limited accretion episodes lasting for $\sim10^7$yr. Many MBH binaries inevitably form during such process, which are now hunted by several observational campaigns, and will be in the future primary targets of GW probes. At the small end of the mass function, MBHs populating spirals and dwarf galaxies evolved in a very quiet environment, without possibly even experiencing a single merger event. Their mass growth is thought to be driven by short accretion episodes following tidal disruption of stars, capture of molecular clouds, and inspirals of compact objects. Different growth modes are expected to result in different spins of the growing objects. Although coherent prolonged accretion via a radiatively efficient thin disk effectively spins up the accreting objects, both MBH binary mergers and accretion of incoherent packets of matter with randomly oriented angular momenta are expected to produce a spin-down. It is therefore likely that the MBH spin distribution reflects their accretion history, being linked with the morphology of their host galaxies. Current measurement, however, are still too sparse and uncertain to test theoretical hypothesis. Future high redshift observations, together with the detection of smaller and smaller MBHs in local galaxies will help in shedding further light on the cosmic evolution of these fascinating objects. In particular, forthcoming GW observations will make possible mass and spin measurements of coalescing MBH binaries out to $z\approx15$ with unprecedented precision, offering a unique opportunity to study their growth along the cosmic history from the very beginning.     

\section*{Acknowledgments}
I am indebted to Marta Volonteri for suggesting a lot of literature relevant to this review, for making figure \ref{figseeds} available and for the modifications to figure \ref{fig1}. I thank Rennan Barkana for the original version of figure \ref{fig1}, and Massimo Dotti for his detailed and thoughtful comments to the manuscript.

\end{document}